\documentclass[journal]{IEEEtran}

\ifCLASSINFOpdf
\else
   \usepackage[dvips]{graphicx}
\fi
\usepackage{url}

\usepackage{graphicx}
\usepackage{amssymb} 
\usepackage{epstopdf}
\usepackage[ruled,norelsize]{algorithm2e}
\makeatletter
\newcommand{\removelatexerror}{\let\@latex@error\@gobble}
\makeatother
\usepackage{pseudocode}
\usepackage{amsthm}
\usepackage{amsmath}
\usepackage{cite}
\usepackage{tikz}
\usepackage{marginnote}
\usepackage{subfigure}
\usepackage{hyperref}
\usepackage{mathtools}
\usepackage{floatrow}
\usepackage{graphicx}
\usepackage{colortbl}
\usepackage{xcolor}
\makeatletter
\newcommand*\bigcdot{\mathpalette\bigcdot@{.65}}
\newcommand*\bigcdot@[2]{\mathbin{\vcenter{\hbox{\scalebox{#2}{$\m@th#1\bullet$}}}}}
\makeatother

\usepackage{booktabs,array}

\newcount\rowc

\makeatletter
\def\ttabular{%
	\hbox\bgroup
	\let\\\cr
	\def\rulea{\ifnum\rowc=\@ne \hrule height 1.3pt \fi}
	\def\ruleb{
		\ifnum\rowc=1\hrule height 1.3pt \else
		\ifnum\rowc=6\hrule height \heavyrulewidth 
		\else \hrule height \lightrulewidth\fi\fi}
	\valign\bgroup
	\global\rowc\@ne
	\rulea
	\hbox to 10em{\strut \hfill##\hfill}%
	\ruleb
	&&%
	\global\advance\rowc\@ne
	\hbox to 10em{\strut\hfill##\hfill}%
	\ruleb
	\cr}
\def\endttabular{%
	\crcr\egroup\egroup}

\begin{document}

\title{Three-Dimensional MRI Reconstruction with 3D Gaussian Representations: Tackling the Undersampling Problem}

\author{Tengya Peng, Ruyi Zha, Zhen Li, Xiaofeng Liu, Qing Zou
\thanks{Tengya Peng is with the Department of Biomedical Engineering , University of Texas Southwestern Medical Center, Dallas, USA. Ruyi Zha is with the Australian National University, Canberra, Australia. Zhen Li and Xiaofeng Liu are with the Department of Radiology and Biomedical Imaging, Yale University, New Haven, USA. Qing Zou is with the Department of Pediatrics, University of Texas Southwestern Medical Center, Dallas, USA (e-mail: Qing.Zou@UTSouthwestern.edu).}}

\maketitle

\begin{abstract}
Three-Dimensional Gaussian representation (3DGS) has shown substantial promise in the field of computer vision, but remains unexplored in the field of magnetic resonance imaging (MRI). This study explores its potential for the reconstruction of isotropic resolution 3D MRI from undersampled k-space data. We introduce a novel framework termed 3D Gaussian MRI (3DGSMR), which employs 3D Gaussian distributions as an explicit representation for MR volumes. Experimental evaluations indicate that this method can effectively reconstruct voxelized MR images, achieving a quality on par with that of well-established 3D MRI reconstruction techniques found in the literature. Notably, the 3DGSMR scheme operates under a self-supervised framework, obviating the need for extensive training datasets or prior model training. This approach introduces significant innovations to the domain, notably the adaptation of 3DGS to MRI reconstruction and the novel application of the existing 3DGS methodology to decompose MR signals, which are presented in a complex-valued format.

\end{abstract}


\IEEEpeerreviewmaketitle

\section{Introduction}

\IEEEPARstart{M}{agnetic} Resonance Imaging (MRI) serves as a critical diagnostic tool in contemporary medicine, offering high-resolution images of the human body without employing ionizing radiation. Three-dimensional (3D) MRI enhances the functionality of conventional two-dimensional (2D) MRI by generating volumetric datasets that allow for manipulation and examination from various angles, thus providing a comprehensive view of human anatomy \cite{subhas2011mri}. Unlike 2D MRI, which produces images in discrete slices, 3D MRI captures extensive data volumes in a single acquisition, often using isotropic voxels to achieve higher spatial resolution with better signal-to-noise ratio. This enhancement facilitates improved detail and clearer distinction of anatomical structures, especially in anatomically complex regions \cite{edelman2009rapid,lusebrink2017t1}. However, 3D MRI is associated with several challenges, including prolonged acquisition time that increases the risk of patient movement, potentially compromising image quality and necessitating repeat scans. Additionally, the extended acquisition time can affect patient comfort, posing ongoing challenges in 3D MRI applications.  
 
Parallel imaging \cite{pruessmann1999sense, griswold2002generalized} is a technique used to decrease the acquisition time of 3D MRI, although it faces certain constraints such as struggling to produce images without artifacts when the undersampling factor is high. Recent developments in compressed sensing \cite{baron2018rapid,feng2017compressed,li2024compressed} offer alternatives to further reduce the acquisition time of 3D MRI while preserving image quality. Techniques such as SPIRiT \cite{lustig2010spirit} and P-LORAKS \cite{haldar2016p}, which leverage iterative self-consistency and low-rank modeling of k-space data, have been shown to enhance the efficiency of parallel imaging, providing complementary methods to accelerate 3D MRI reconstruction. Additionally, restoring MRI from undersampled acquisitions using deep-learning-based algorithms received much attention. For example, unrolled model is a technique that unrolls the iterations of optimization as neural network blocks \cite{huang2023self,xin2024rethinking}. The diffusion model is another scheme that treats MRI undersampling as a process of gradually adding Gaussian noise to simulate the degradation from full to undersampled data and learns to recover the fully-sampled distributions from undersampled ones with noises \cite{liu2023score,korkmaz2023self,bangun2025mri}. While these approaches provide promising results, training data for pretraining is required.  To overcome the limitation of the need for training data for pretraining, patient-specific reconstruction schemes are proposed. These frameworks usually involve exploration in image space by forcing consistency in frequency space with certain constraints such as total-variation (TV) constraints or low-rank plus sparsity constraints. It was found that a randomly initialized convolutional neural network (CNN) can inherently capture low-level image statistics, serving as an effective implicit prior for solving inverse problems and such scheme is usually termed as deep image prior \cite{ulyanov2018deep,liang2025analysis}. This insight enables a randomly initialized convolutional decoder (ConvDecoder) \cite{darestani2021accelerated} to model the MRI image space and recover the images, as the CNN architecture inherently imposes spatial constraints by capturing local pixel dependencies. Implicit neural representation (INR) is another state-of-the-art technique for modeling 3D volumes by mapping continuous spatial coordinates to signal values using powerful spatial encoding. There are two styles for INR: siren-encoding \cite{mildenhall2021nerf}, and hash encoding\cite{muller2022instant,zha2022naf}. Jang et al. \cite{jang2024nerf} employed a hash-encoded INR to address the challenge of undersampled MRI reconstruction slice by slice for the 3D MRI volume using projection slice theorem.

Recently, 3D Gaussian representation (3DGS) has emerged as a notable alternative for INR within the realm of 3D modeling and rendering, offering a versatile and robust framework for efficiently rendering complex scenes. 3DGS models 3D volumes as collections of Gaussian points, enabling the representation of both geometric and visual attributes. The Gaussian mixture model (GMM) is conceptually similar to 3DGS but generates predictions as a combination of multiple Gaussian distributions. In GMM, the probability density is influenced by the distance of a pixel value from a learned mean value. In contrast, 3DGS calculates the probability density based on the spatial distance, and introduces an additional learnable parameter that multiplies with the probability density to produce the final output. Hou et al. \cite{hou2025integrating} integrated GMM into their MRI reconstruction pipeline to remove the artifacts and noise produced in iterative optimization process.

While 3DGS has proven effective in computer vision and certain medical imaging applications, its potential in MRI remains largely unexplored. Several studies \cite{li2023sparse,lin2024learning,cai2025radiative} have explored 3DGS for X-ray image reconstruction. However, these approaches generally fail to establish a direct connection between the Gaussian representation and volumetric (voxel-based) representations due to the lack of an explicit transformation between the two. The work of \cite{zha2024r} is the first to bridge this gap by introducing a CUDA-accelerated approach that enables the efficient transformation of 3D Gaussian points into grid points within a 3D volume. The model retains a tile-based voxelization design for efficiency. In this study, we aim to apply 3DGS to explore the undersampled MRI reconstruction problem within the context of isotropic resolution 3D MRI.

Recent studies have increasingly focused on training 3DGS in the frequency domain, proposing a progressive coarse-to-fine learning paradigm \cite{yu2024mip,jung2024relaxing,seo2024flod,deng2024efficient,chen2025dashgaussian}. Specifically, FreGS \cite{zhang2024fregs} implements this approach by prioritizing low-to-high frequency components, leveraging low-pass and high-pass filters in Fourier space to enhance performance. These studies have inspired our work on using 3DGS for MRI reconstruction at high acceleration factors using a long-axis splitting strategy that we will introduce in the following section. The long-axis splitting strategy starts with only a limited number of Gaussian points to catch the coarse information, and by increasing the number of Gaussian points in a designed splitting pattern, we gradually recover the fine details in the images.

In this study, we introduce a novel approach employing 3D Gaussian representations, termed 3DGSMR, for reconstructing isotropic resolution 3D MRI from undersampled measurements. The 3DGSMR model is initialized with the inverse Fast Fourier Transform (iFFT) applied to undersampled k-space data, which serves as the foundation for subsequent optimization aiming at detailed image recovery. The voxelized 3D MR volume is derived from the 3D Gaussian representation. To maintain data consistency, we implement data-consistency constraint in k-space to ensure that the Fourier transform of the voxelized volume aligns with the undersampled k-space data. 3DGSMR has demonstrated exceptional performance in the task of undersampled 3D MRI reconstruction. It effectively decomposes the complex-valued MRI data into 3DGS. Additionally, we have refined the densification strategies for 3DGS within this work, enhancing the overall performance of the reconstruction for high acceleration factors.

The contributions and innovations of this research include: 1) The usage of a novel 3D MRI representation model. In this model, each voxel of the 3D MRI is represented by a combination of learnable 3D Gaussian points. This innovative approach facilitates a continuous representation of the image, enhancing the granularity and flexibility of the imaging process; 2) The expansion of the existing 3DGS framework to enable the decomposition of complex-valued 3D volumes. This extension broadens the applicability of 3DGS, allowing for more sophisticated manipulation and analysis of MRI data; 3) A new densification strategy that improves the reconstruction quality at high acceleration factors.

\section{Method}

\subsection{3D MRI reconstruction from undersampled data}

The main focus of this work is to recover the 3D MR images $\mathbf X\in\mathbb{C}^3$ from undersampled multichannel k-space measurements. Mathematically speaking, images are acquired by multichannel measurement in k-space:
\begin{equation}\label{key}
\mathbf b = \mathcal A(\mathbf X) + \mathbf n,
\end{equation}
where $\mathbf n$ is zero mean Gaussian noise matrix that corrupts the measurements during the acquisition, and $\mathcal{A}$ is the forward operator consisting of coil sensitivity maps, undersampling mask and the FFT.

To reconstruct $\mathbf X$ from the undersampled measurements $\mathbf b$, the following minimization problem needs to be solved:
\begin{equation}\label{mini}
\mathbf{X}^* = \arg\min_{\mathbf{X}\in\mathbb{C}^3}||\mathcal A(\mathbf X)-\mathbf{b}||^2 + \lambda\cdot\mathcal{R}(\mathbf{X}),
\end{equation}
where $||\mathcal A(\mathbf X)-\mathbf{b}||^2$ is the data-consistency term and $\mathcal{R}(\mathbf{X})$ is a regularization term to penalize the solution. $\lambda$ is the balancing parameter that balances the data-consistency term and the regularization term.

\subsection{3D Gaussian representation model for 3D MRI}

In this work, we model each voxel in the 3D MRI volume $\mathbf{X} = [x_j]$ as the combination of a set of learnable 3D Gaussian points. Here, $\mathbf{x_j}\in\mathbb{C}$ is the element at spatial location $\mathbf{j}=(x_j,y_j,z_j)$. With this, we can write
\begin{equation}\label{image_model}
x_j = \sum_{i=1}^{M}G_i^3(\mathbf{j} | \rho_i, \mathbf{p}_i, \mathbf{\Sigma}_i),
\end{equation}
where $G_i^3\in\mathbb{G}^3$ is a local 3D Gaussian representation component that defines a local Gaussian distribution density field. It is formulated as 
\begin{equation}\label{key}
G_i^3(\mathbf{j} | \rho_i, \mathbf{p}_i, \mathbf{\Sigma}_i) = 
\rho_i\cdot\exp\left(-\frac{1}{2}(\mathbf{j}-\mathbf{p}_i)^T\mathbf{\Sigma_{i}}^{-1}(\mathbf{j}-\mathbf{p}_i)\right),
\end{equation}
where $\rho_i\in\mathbb{C}$, $\mathbf{p}_i\in\mathbb{R}^3$, and $\mathbf{\Sigma_i}\in\mathbb{R}^{3\times 3}$ are learnable parameters representing the central density values, the position of the center, and the covariance matrix in the Gaussian point, respectively. Note that $\rho_i$ is in the complex domain to match the complex-valued form of the MR images in our task. $\mathbf{p}_i$ defines the position of the Gaussian point cloud, while $\mathbf{\Sigma}_i^{-1}$ defines the shape size and orientation. $\mathbf{\Sigma_i}$ can further be decomposed into rotation and scaling as 
\[\mathbf{\Sigma}_i = \mathbf{R}_i\mathbf{S}_i\mathbf{S}_i^T\mathbf{R}_i^T,\]
where $\mathbf{R}_i$ and $\mathbf{S}_i$ are the rotation and scaling matrix. The rotation matrix is constructed using quaternions with 4 parameters governing the rotation of the Gaussian point. These parameters collectively describe the local 3D Gaussian representation derived from a normal 3D Gaussian distribution. 

The optimization process of 3DGSMR can be viewed as exploring various combinations of Gaussian point clouds, accounting for different scales, rotations and complex values, to accurately model the MRI volume. This process is guided by loss functions, typically comprising L1 or L2 norms alongside regularizations, which serve as constraints to direct the optimization efforts. Several factors influence the training process including the method of initialization, adaptive density control \cite{hernandez1990density}, and the implementation of constraints or regularization techniques.

\subsection{Initialization of 3D Gaussian representation}

Initialization plays a vital role in the training of 3D Gaussian framework due to the vast range of diverse possible combinations of Gaussian point clouds. Usually, a coarse and primitive reconstruction result is used to guide the 3DGS initialization, from which the model can evolve. In the MRI setting, the iFFT of undersampled k-sapce data is used as the initialization in this work, as iFFT from undersampled k-sapce data is freely accessible in the setting of MRI reconstruction. In the iFFT reconstruction, we randomly sample $M$ grid points as the center of the initial Gaussian points. The quantity of initial Gaussian points plays a critical role in shaping the performance of the model. Then we set the scales of the initial Gaussian points as the average of the distances to the three nearest grid points and assume no rotations. For the central density values of the Gaussian points for both the imaginary and real parts, they are initialized by scaling down with factor $k$ from the values of the sampled grid points.

\subsection{Adaptive control}

\begin{figure}[htbp!]
\centering
   \includegraphics[width=0.8\textwidth]{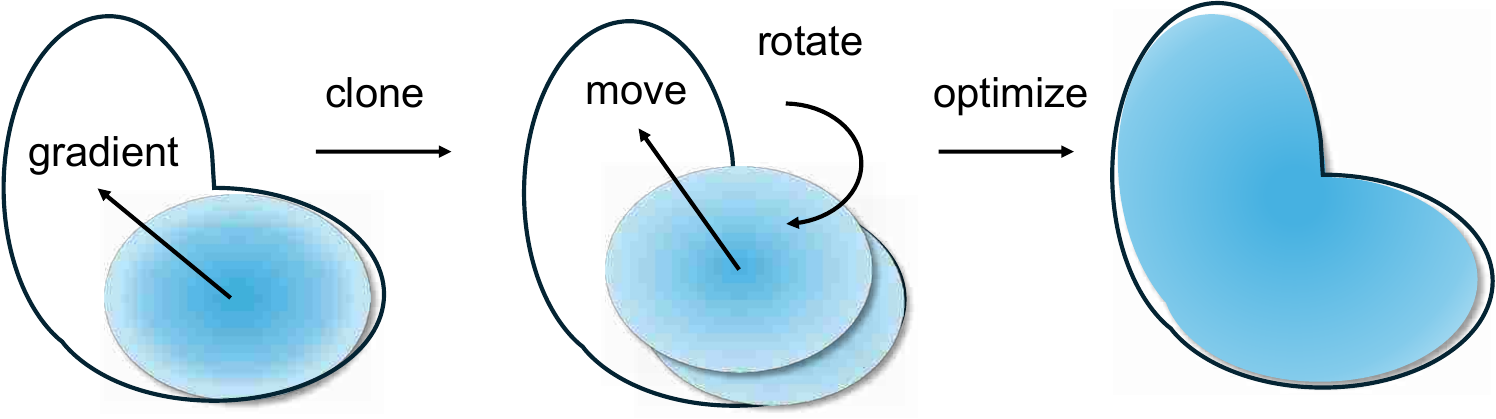}
  \caption{Cloning method in adaptive control. The cloned Gaussian point will be further optimized to fit into unreconstructed region.}
  \label{cloning_fig}
\end{figure}

Adaptive control functions as a mechanism for densifying Gaussian points when the current number of Gaussian points is insufficient to accurately represent the MR volume. It is triggered when the normalized gradients of the Gaussian centers exceed a set threshold $\tau$. In such cases, cloning or splitting is needed based on the current sizes of Gaussian points. If the sizes of the current Gaussian points exceed a threshold $s$, splitting will be activated; otherwise, cloning will be triggered. For cloning, another Gaussian point of the same shape and scale is added in the same position. The central density values are scaled to half. The newly cloned Gaussian points will be trained to fill the empty spaces that are under-reconstructed. Such process is illustrated in Figure \ref{cloning_fig}. 

\begin{figure}[htbp!]
\centering
   \includegraphics[width=0.9\textwidth]{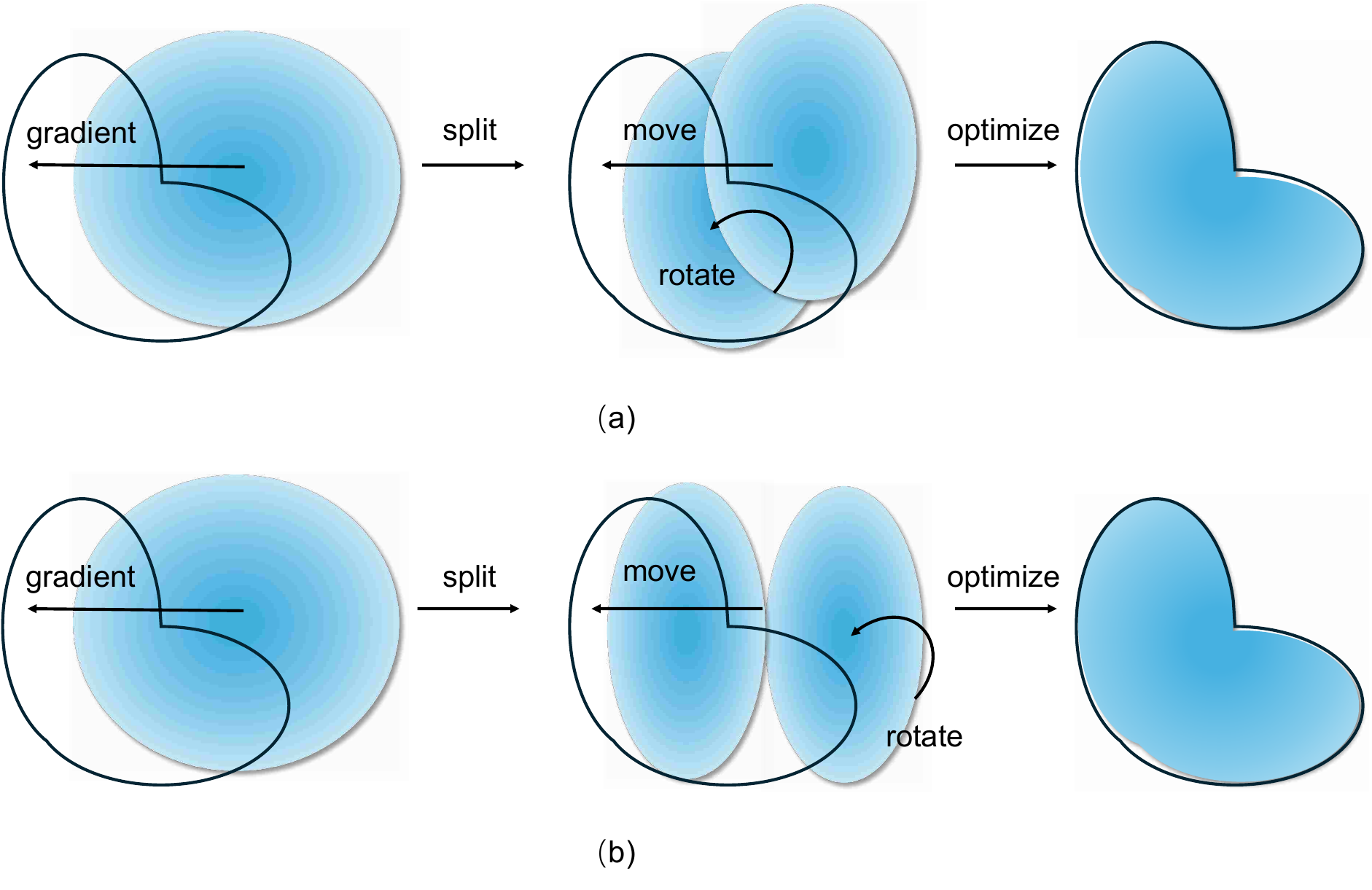}
  \caption{(a) Original splitting method in adaptive control. (b) The long-axis splitting approach. A large Gaussian point splits into two small identical Gaussian points along the longest axis which are then further optimized to fit into unreconstructed region.}
  \label{splitting_fig}
\end{figure}

In terms of the splitting, the Gaussian point is split into two identical Gaussian points, each with size equals to the original Gaussian size scaled by a factor of 1/1.6, and the central values are scaled down by half. These setting are based on the design of primitive 3DGS \cite{kerbl20233d}. Unlike cloning, the positions of the two newly split  Gaussian points follow a normal distribution centered around the previous center position. This results in two split Gaussian points having different proportions of overlaps. This process is illustrated in Figure \ref{splitting_fig} (a). Deng et al. \cite{deng2024efficient} indicated that excessive overlaps not only increase computational costs but also introduce blurring artifacts, which obscure fine details and hinder the accuracy of scene reconstruction. To overcome this problem, a long-axis splitting approach is proposed. This splitting approach removes cloning and splits the Gaussian points along the longest axis to half, while the other two axes are scaled by a factor of 0.85 of the original size. The density values of each child Gaussian point are scaled to 0.6 of original values. Based on \cite{deng2024efficient}, such setting ensures that the two Gaussian points after splitting do not overlap on the longest axis, ensuring consistent overall shape before and after the splitting. Such concept is depicted in Figure \ref{splitting_fig} (b). We show in this work that the long-axis splitting approach improves the reconstruction of high-frequency details and improves the reconstruction results for high acceleration factors.

To improve efficiency, we prune Gaussian points whose complex central-density magnitude falls below a threshold yet exhibits large loss gradients during the process of adaptive control.

\begin{figure*}[htbp!]
\centering
\includegraphics[height=290pt,width=470pt]{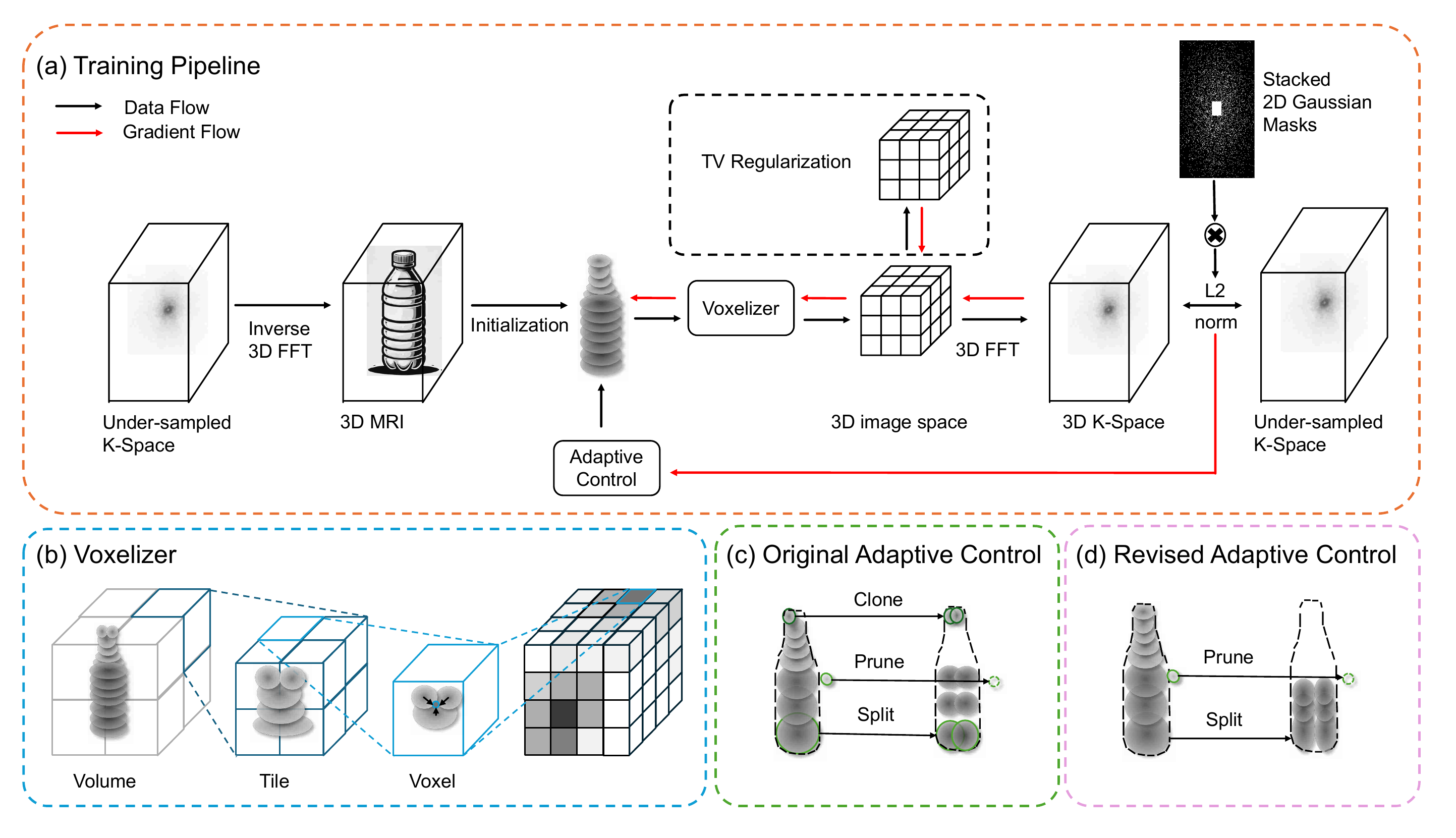}
\caption{Training pipeline and components of 3DGSMR. (a) End-to-end pipeline showing data and gradient flow from initialization to output, comprising initialization, voxelization, TV regularization, Fourier transform, and adaptive control. (b) Voxelizer schematic: sparse 3DGS points are aggregated onto grid locations via weighted summation to form a voxelized volume. (c) Original 3DGS adaptive control: cloning (duplicate at the same location), splitting (one Gaussian into two with positions drawn from normal distributions), and pruning (removing low-amplitude Gaussians). (d) Our modified adaptive control for high acceleration factors: cloning is removed; splitting proceeds along the longest axis without overlap, preserving the original shape.}
\label{illu_figure}
\end{figure*}

\subsection{Voxelization}
Voxelization serves as a crucial bridge for transforming 3DGS into voxelized volumetric representations. This technique represents one of the key innovations introduced in \cite{zha2024r}. Our model builds upon such voxelization pipeline, which originates from the rasterization process used in the rendering part of the primitive 3DGS model \cite{kerbl20233d}. In this process, neighboring Gaussian points are aggregated to compute the central intensity values of voxels. The core of this voxelization pipeline is the tile-based design, which divides the image volume into multiple non-overlapping patches. Each tile contains a subset of the voxels. Although the Gaussian distribution is continuous and theoretically infinite, we limit its influence to within three standard deviations from its center, following the three-sigma rule \cite{threesigma}. The radius is chosen based on the largest standard deviation along any axis of the Gaussian point. A Gaussian point is considered to intersect a tile if its 3$\sigma$ bounding box overlaps the tile's spatial extent. The pipeline will first identify the Gaussian points that fall into the tile and emit tile-Gaussian pairs (tile's ID and Gaussian's ID). Then the pairs are sorted by the tiles' IDs. Because each tile’s pairs are contiguous at this moment, we can record a starting and ending index for every tile and therefore gather all the Gaussian points allocated in the tiles backwards. The tile design offers two primary benefits: efficient on-chip data reuse and parallel execution. For each tile, the kernel caches its Gaussian in shared memory --- streaming them in batches when they exceed on-chip capacity—while tile-mapped threads iterate over the staged set to accumulate contributions at voxel centers. The voxelization pipeline is depicted in Fig. \ref{illu_figure} (b). For a more detailed analysis and explanation of the tile-based design, we would refer the readers to this survey \cite{chen2025survey3dgaussiansplatting}.

\subsection{Complexity and GPU memory analysis}
Compared to voxel-based representations, 3DGS offers a more compact encoding in terms of spatial complexity. In the proposed model to represent MRI volume, each Gaussian point typically requires twelve parameters to store (3 for center location, 3 for scaling, 4 for rotation quaternions, and 2 for real and imaginary part of the complex density value), and significantly fewer Gaussian points are needed to represent a volume than in voxelization. For instance, in our model, we limit the number of Gaussian points to no more than 400,000 for a volume of size $400\times303\times165$. This yields 
$400,000\times12=4,800,000$ parameters — far fewer than the 
$400\times303\times165\times2=39,996,000$ parameters required in a dense voxel grid representation, which demonstrates a more memory-efficient
image representation.

Due to the introduction of tile-based design, the time complexity of 3DGS has been reduced significantly. Suppose an image volume with $M=Z*H*W$ voxels and such image volume is partitioned into tiles of $N$ voxels in dimension $N = Nx*Ny*Nz$. Then the total number of tiles is $T=[Z/Nx]*[H/Ny]*[W/Nz]$, where the notion [] represents the rounding up operator. Let $Q$ be the number of Gaussian points assigned to the busiest tile. With shared memory processing in batches of size $N$, such tile needs $[Q/N]$ rounds. Because tiles run in parallel, the overall step time is dominated by this tile with time complexity proportional to $[Q/N]*N$. Thus the complexity scales with the maximum per-tile load $Q$. Poor initialization or training can leave Gaussian points so large that they span multiple tiles, which substantially increases the time complexity. In the MRI setting in this work, it is found that Gaussian points tend to shrink during training to capture high-resolution details, and most often end up covering only one or two tiles. Fourier transform is applied to the voxelized volume to compute the loss after voxelization.

In terms of GPU memory, each Gaussian has 12 learnable parameters: center location, scaling, and rotation quaternions. From rotation quaternions and scaling, 3×3 covariance matrix $\Sigma$ and its inverse $\Sigma^{-1}$ are computed (both symmetric, thus 6 scalars each). Moreover, storing Gaussian–tile pair IDs incurs memory proportional to the number of pairs. The per-tile shared buffer holds tile-sized batches of values — 6 scalars for $\Sigma^{-1}$, 3 for the Gaussian center, and 2 for the complex MRI value of a Gaussian, for every tile group. Image volume and k-space volume each require $Z*H*W*2$ parameters. During backpropagation, tensors of comparable size are instantiated for gradients. Additional temporaries arise from matrix operations such as multiplication, computation of Gaussian's contributions to voxels, and FFT steps. In general, the GPU consumption grows with the volume size, number of Gaussians, and how Gaussians intersect with tiles (which is often driven by sizes of Gaussians).

\subsection{Training}

Based on the above described optimization pattern for MRI reconstruction, the following loss function is solved for reconstructing the isotropic resolution 3D MRI:
\begin{eqnarray}\nonumber
\mathcal{L}(\rho_i,\mathbf{p}_i,\mathbf{\Sigma_i}) &=& ||\mathcal A\left([\sum_{i=1}^{M}G_i^3(\mathbf{j} | \rho_i, \mathbf{p}_i, \mathbf{\Sigma}_i)]\right) -\mathbf{b} ||\\\nonumber
&+ & \lambda\cdot TV\left(|\sum_{i=1}^{M}G_i^3(\mathbf{j} | \rho_i, \mathbf{p}_i, \mathbf{\Sigma}_i)|\right). 
\end{eqnarray}
Note that the TV loss was applied to the magnitude of the 3D MR volumes. The overall training scheme is depicted in Figure \ref{illu_figure}.

In the setting of our training process, the maximum number of Gaussian points is set to 400 thousand (400k) to stop further densification. The densification is done in every 100 iterations.

\section{Experiments}

\subsection{Dataset}
This research study was conducted using data acquired from human subjects. Seven fully-sampled k-space sets acquired from seven healthy volunteers were used in this work to simulate undersampling scenario of different acceleration factors. The Institutional Review Board at the University of Texas Southwestern Medical Center approved the acquisition of the data, and written consents were obtained from all subjects. The data was acquired on a 1.5T scanner (Ingenia, Philips Healthcare, Best, Netherlands) with a 20-array dStream Head Neck coil. The dataset was acquired in a fully-sampled fashion using a single-shot T1-weighted inversion recovery turbo field echo (TFE) sequence. The sequence parameters include: field of view = 350 mm $\times$ 250 mm $\times$ 200 mm corresponding to Foot-Head direction $\times$ Anterior-Posterior direction $\times$ Right-Left direction, TE/TR = 3.4/7.2 ms, flip angle = 10$^\circ$, inversion time (TI) = 1122 ms, TFE factor = 303, and shot interval = 2300 ms. The acquisition time for each subject is 6 minutes and 24 seconds.

We used an automatic algorithm implemented on the scanner to pre-select the best coils, that provide the best signal-to-noise-ratio in the region of interest. A PCA-based coil combination using singular value decomposition is used such that the approximation error is $<$ 5\%.

To simulate clinical scenarios, we applied 3D masks built based on stacked of 2D masks sampled from a 2D Gaussian distribution, and such masks enabled undersampling in phase-encoding direction and the readout direction is fully sampled. Furthermore, the masks have a fully sampled center calibration region to ensure better incorporation of the low-resolution regions of MRI images. To strike a balance between acceleration and performance, our experiments tested acceleration factors of 2, 4, 6, 8, 10, 12, 14, and 16 respectively.

\subsection{Implementation details}

The proposed 3DGSMR scheme was implemented based on the PyTorch library and realized using a single NVIDIA A6000 graphic card that provides 48 GB of GPU memory. The model was initialized with a fixed number of Gaussian points based on the iFFT images obtained from the undersampled k-space. These initial Gaussian points were then trained based on the loss function and densified by cloning and splitting until the maximum allowable number of Gaussian points was reached. Three quantitative metrics, structural similarity index measure (SSIM) \cite{wang2004image}, peak signal-to-noise-ratio (PSNR) \cite{wang2004image}, and Learned Perceptual Image Patch Similarity (LPIPS) \cite{zhang2018unreasonable} were calculated to assess the reconstruction performance and determine the optimal stopping criteria for training. SSIM evaluates structural similarity between images, focusing on luminance, contrast, and texture. PSNR measures pixel-level differences, assessing noise and distortion, but lacks perceptual sensitivity. LPIPS captures perceptual similarity by comparing high-level features with deep neural networks, reflecting human visual perception of textures, shapes, and image quality.

The training process was progressive, continuing until the SSIM, PSNR, and LPIPS metrics reached a plateau, meaning that progressive training would bring little or even no more improvement.

\section{Results}

\subsection{Ablation studies}

In this section, we detail the process of the determination of the hyper-parameters in the proposed framework. Data from one subject was used to identify the optimal hyper-parameters in the proposed scheme. We then used the hyper-parameters to generate the experimental results across the whole dataset reported in this paper. 

We first conducted the ablation study on the number of of initial Gaussian points $M$, scaling factor $k$, threshold $s$ between splitting and cloning, and the coefficient of the TV loss $\lambda$. The study was conducted on a single subject's data with an acceleration factor of 8, and the training was performed with the same number of iterations. The results are shown in Table \ref{tab:ablation_merged}. The highlighted parameters in the table are the ones that we picked in the proposed framework. In summary, we selected $M=200$k, $k=0.2$, $s=0.01$, and $\lambda=0.1$ to achieve an optimal solution that balances between performance and training efficiency.

\begin{table}[ht]
\centering
\begin{tabular}{|l|c|c|c|c|}
\hline
\textbf{ } & \textbf{SSIM}$\uparrow$ & \textbf{PSNR}$\uparrow$ & \textbf{LPIPS}$\downarrow$ & \textbf{Time}$\downarrow$ \\
\hline
$M=$ 50k & 0.936 & 32.47 & 0.111 & 11min27s \\
$M=$ 100k & 0.938 & 32.13 & 0.105 & 9min36s \\
$M=$ 200k \cellcolor{red!50} & 0.941\cellcolor{red!50} & 31.76 & 0.097\cellcolor{red!50} & 9min40s \\
$M=$ 400k & 0.935 & 32.05 & 0.106 & 11min36s \\
\hline
$k=0.1$ & 0.941 & 31.90 & 0.097 & 9min33s \\
$k=0.15$ & 0.940 & 32.34 & 0.096 & 9min30s \\
$k=0.2$ \cellcolor{red!50}& 0.941\cellcolor{red!50} & 32.19 \cellcolor{red!50}& 0.096 \cellcolor{red!50}& 9min33s \\
$k=0.5$ & 0.940 & 32.48 & 0.101 & 9min28s \\
\hline
$s=0.01$\cellcolor{red!50} & 0.941 \cellcolor{red!50} & 32.50 & 0.097\cellcolor{red!50} & 9min37s \\
$s=0.02$ & 0.941 & 32.60 & 0.103 & 14min21s \\
$s=0.05$ & 0.941 & 32.76 & 0.103 & 14min46s \\
$s=0.1$ & 0.941 & 32.41 & 0.100 & 17min2s \\
\hline
$\lambda=0$ & 0.939 & 31.68 & 0.097 & 9min36s \\
$\lambda=0.1$\cellcolor{red!50} & 0.942 \cellcolor{red!50} & 32.87 \cellcolor{red!50}& 0.095\cellcolor{red!50} & 9min38s \\
$\lambda=0.5$ & 0.941 & 32.21 & 0.097 & 9min37s \\
$\lambda=1$ & 0.940 & 31.83 & 0.097 & 9min34s \\
$\lambda=100$ & 0.941 & 32.06 & 0.097 & 9min31s \\
\hline
\end{tabular}
\caption{Ablation study results with optimal choices highlighted.}
\label{tab:ablation_merged}
\end{table}

\begin{figure}[htbp!]
\centering
   \includegraphics[width=1\textwidth]{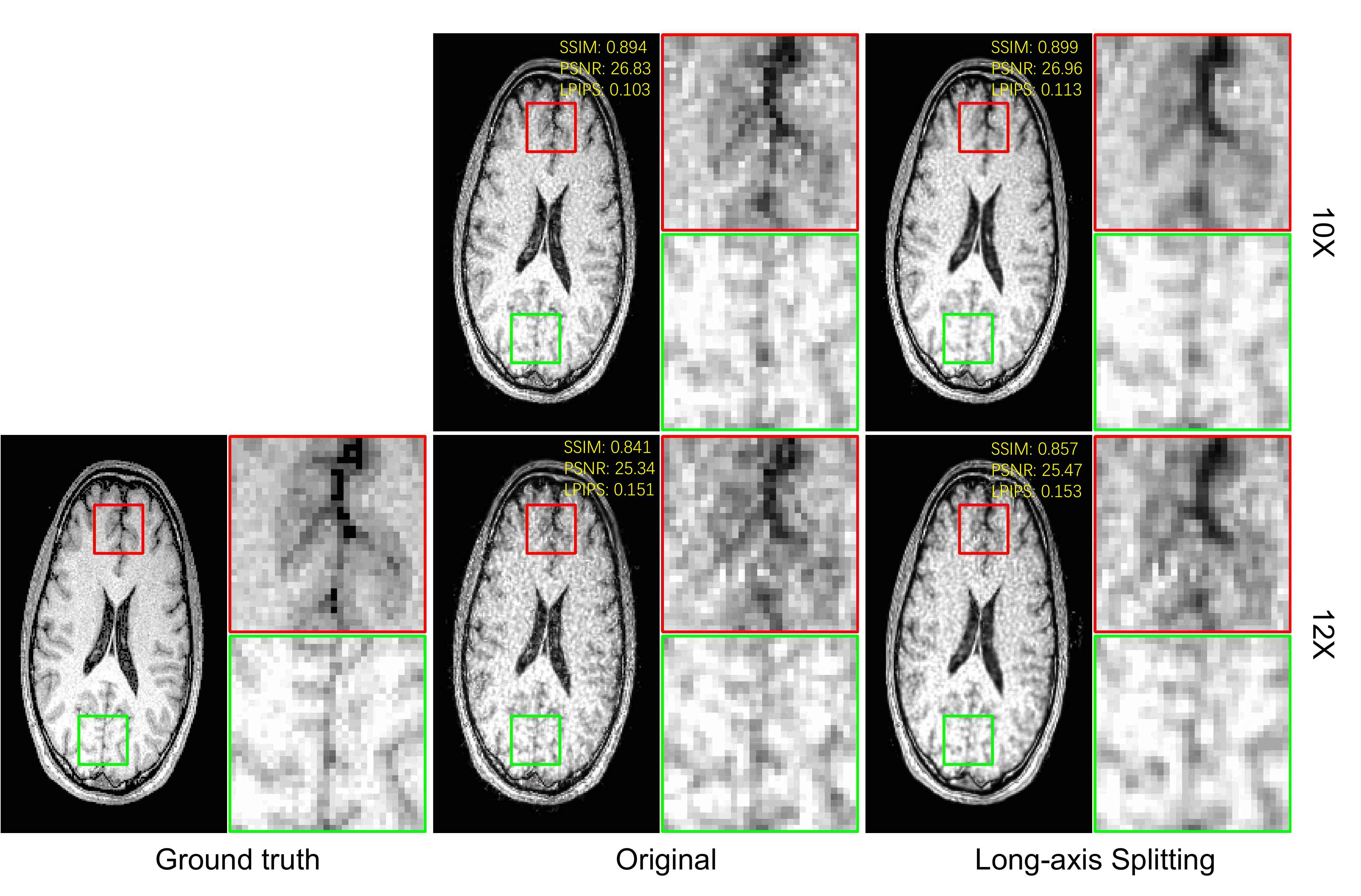}
  \caption{Visual comparison with SSIM, PSNR, and LPIPS of the reconstructed results from the original optimization method versus the new optimization method that introduces long-axis splitting and abandoning cloning along with initializing using only 500 Gaussian points.}
  \label{fig:longaxis}
\end{figure}

\begin{table}[htbp!]
\centering
\begin{tabular}{|p{2.8cm}|c|c|c|c|}
\hline
 & \textbf{SSIM}$\uparrow$ & \textbf{PSNR}$\uparrow$ & \textbf{LPIPS}$\downarrow$ & \textbf{Time}$\downarrow$ \\ \hline
original cloning \& original splitting \& $M=200000$ & 0.911 & 30.28 & 0.131 & 9min36s \\ \hline
original cloning \& original splitting \& $M=500$ & 0.804 & 29.02 & 0.268 & 24min3s \\ \hline
original cloning \& long-axis splitting \& $M=500$ & 0.912 & 30.53 & 0.144 & 17min33s \\ \hline
original splitting \& $M=500$ & 0.794 & 28.95 & 0.276 & 24min35s \\ \hline
long-axis splitting \& $M=200000$ & 0.915 & 30.17 & 0.125 & 10min31s \\ \hline
long-axis splitting \& $M=500$ \cellcolor{red!50}& 0.919\cellcolor{red!50} & 30.94\cellcolor{red!50} & 0.121\cellcolor{red!50} & 17min22s \\ \hline
\end{tabular}
\caption{Quantitative results for different adaptive control strategies on one subject with an acceleration factor of 10.}
\label{tab:longaxis}
\end{table}

\begin{figure*}[htbp!]
\centering
   \includegraphics[width=\textwidth]{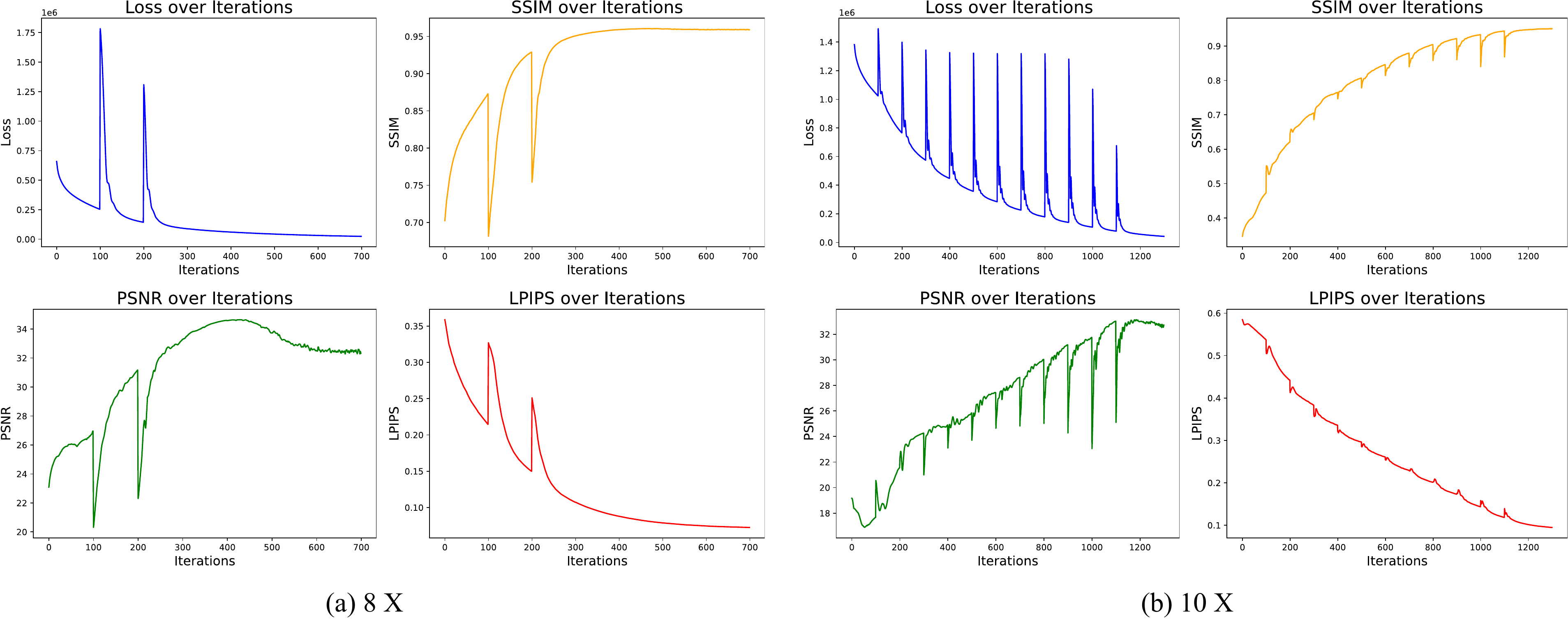}
  \caption{The progression of reconstruction loss and evaluation metrics (SSIM, PSNR, and LPIPS) is shown for one subject, with an acceleration factor of 8 for the original optimization method and 10 for the long-axis splitting strategy under high acceleration. (a) Curves corresponding to an acceleration factor of 8. (b) Curves corresponding to an acceleration factor of 10.}
  \label{loss_curve}
\end{figure*}

\subsection{Splitting strategy for high acceleration factors}

While with the original splitting strategy shown in Figure \ref{splitting_fig} (a), we noticed that the reconstruction results degraded significantly with blurring artifacts when the acceleration factor is high (e.g. 10-fold or 12-fold undersampling). To improve the reconstruction performance, a different splitting strategy is studied, inspired by the work of \cite{deng2024efficient}. Specifically, we modified the original adaptive control by abandoning cloning, and applying long-axis splitting as shown in Figure \ref{splitting_fig} (b), and initializing with less number of Gaussian points (e.g. $M=500$). We investigated such strategy based on data of one subject with acceleration factor 10. The quantitative results obtained from different adaptive control strategies on this subject are shown in Table \ref{tab:longaxis}. It is suggested that initializing with 500 Gaussian points, using the long-axis splitting strategy, and removing cloning outperforms other combinations, indicating this modified optimization strategy offers advantages at high acceleration factors. To further compare the reconstruction results for high acceleration factors, a visual comparisons between the original splitting strategy and the long-axis splitting strategy was shown in Figure \ref{fig:longaxis}. The results of the original splitting and cloning method show more noticeable distortions and blurring, whereas the optimization with long-axis splitting reduces these artifacts.

It is also worth mentioning here that for the low-to-mid acceleration factors (e.g. 2-fold, 4-fold, 6-fold, and 8-fold), different strategies used in Table \ref{tab:longaxis} play no role on the reconstruction performance, indicating that the long-axis splitting strategy works only in the cases that the acceleration factors are high.

\begin{figure*}[htbp!]
\centering
   \includegraphics[width=0.8\textwidth]{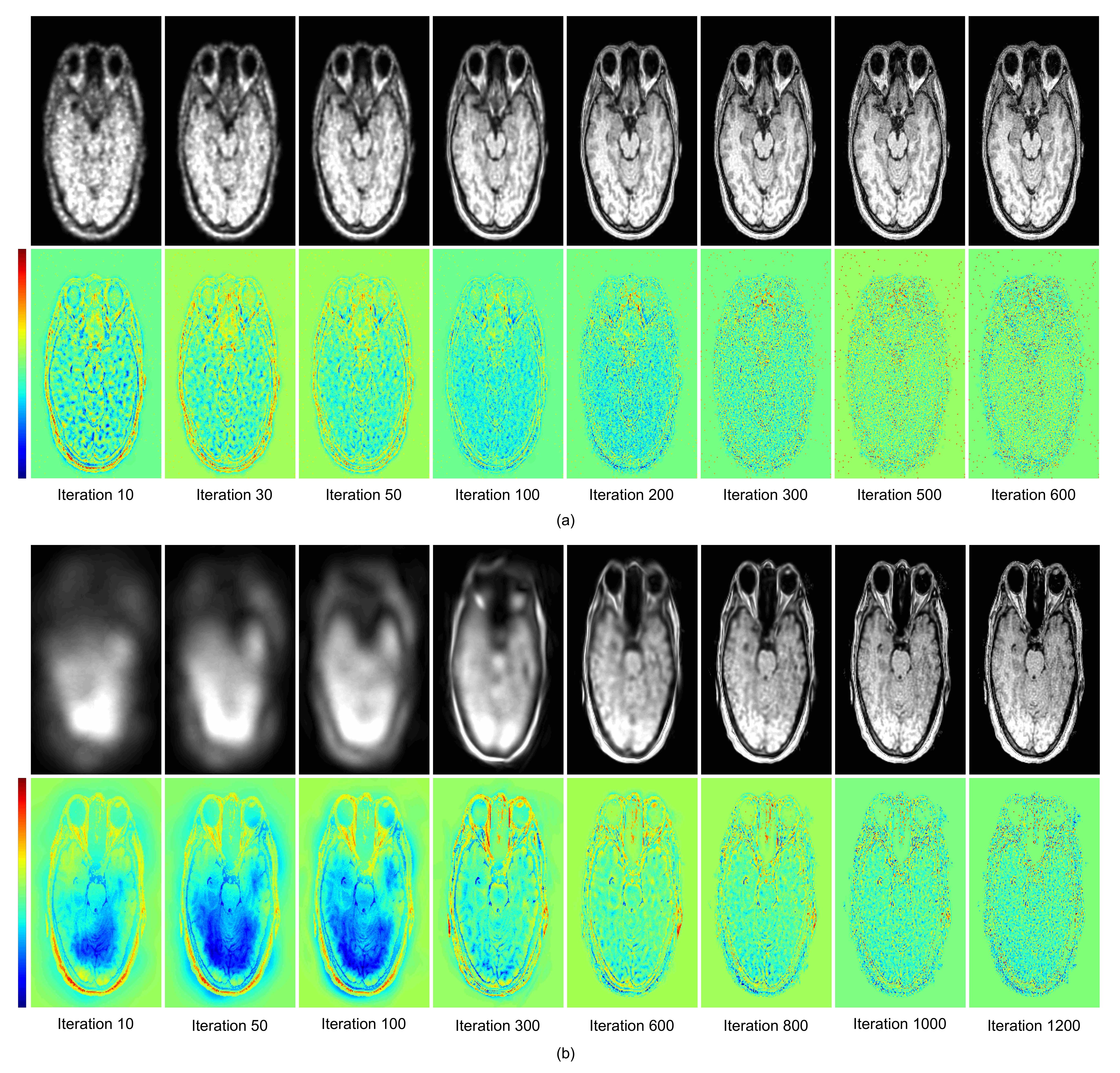}
  \caption{Illustration of the convergence progress. (a) Acceleration factor of 8. The proposed method reaches acceptable results after 200 iterations. In subsequent iterations, additional details are recovered, leading to improved image quality. (b) Acceleration factor of 10 with long-axis splitting. Long-axis splitting helps recover fine details at higher acceleration factors.}
  \label{mid_recons}
\end{figure*}

\subsection{Stopping criterion and reconstruction time}

The experiments in this work are built on the premise that we begin with fully sampled images and manually apply undersampling masks to simulate undersampling scenarios. The reconstruction process is done without the usage of the fully sampled data anymore and only the quantitative results' computation involves the usage of fully sampled data. During the reconstruction process, important metrics including SSIM, PSNR, and LPIPS were recorded for each iteration using TensorBoard. The plots of these metrics against the number of iterations are shown in Figure \ref{loss_curve}. Since 3DGSMR can be trained iteratively and continuously, we implement a stopping criterion to halt training when SSIM, PSNR, and LPIPS plateaued as the reconstruction progressed. We first conduct such experiments to identify the optimal stopping point and hyper-parameters for our model, which are then followed during later deployment.  

In Figure \ref{mid_recons}, the intermediate reconstructions from one subject using acceleration factors of 8 and 10 based on the original cloning and splitting and the long-axis splitting are shown, depicting how the image quality improves visually with the increase of the number of iterations.

\begin{figure}[htbp!]
\centering
   \includegraphics[width=0.95\textwidth]{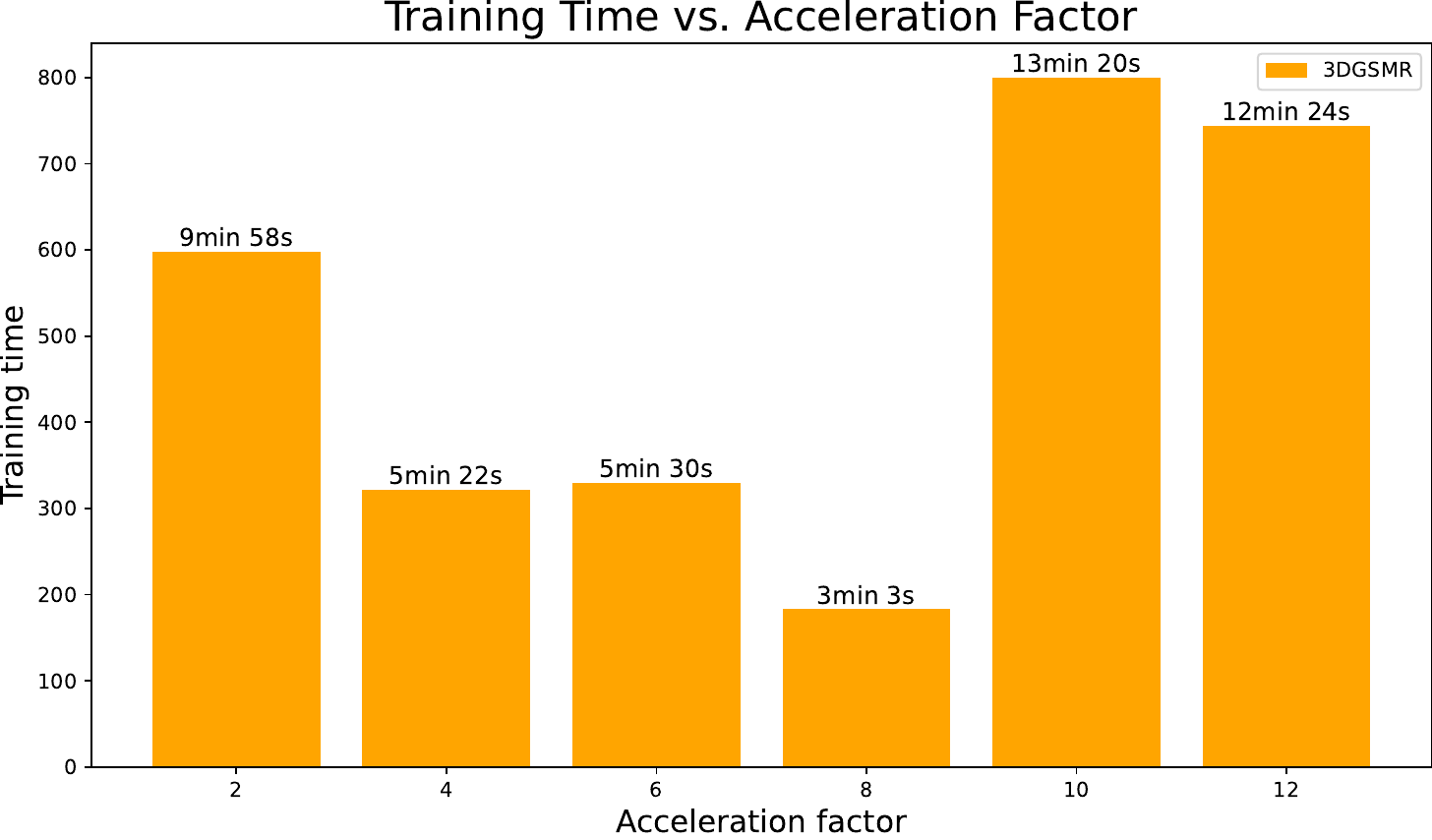}
  \caption{Average reconstruction time for all the dataset across acceleration factors of 2, 3, 4, 6, and 8.}
  \label{recon_time}
\end{figure}

Figure \ref{recon_time} illustrates the average reconstruction time for all subjects across various acceleration factors. For acceleration factors of 10 and 12, long-axis splitting is applied. Note that the time reported in the ablation study includes per-iteration evaluation-metric computation and tensorBoard loss logging. In contrast, the reported reconstruction time is measured separately, excluding these overheads. Under a fixed optimization method, we observe shorter runtime at higher acceleration factors. This trend may arise from the reduced data-consistency constraints, which facilitate faster convergence. A rule-of-thumb estimate is 0.53 seconds per iteration. Regarding the time transforming 3DGS representation into volume representation, it only costs an average 2 milliseconds.

The peak GPU memory during training remained below 10 GB --- typically around 5 GB --- for both the original desnisfication strategy at low–mid acceleration and the long-axis–splitting strategy at high acceleration for a volume of size $400\times303\times165$. During inference, voxelizing the 3D Gaussian representation of MR to volume representation uses approximately 2.36 GB of GPU memory.

\subsection{Effectiveness of the proposed framework}

\begin{figure*}[htbp!]
\centering
   \includegraphics[width=\textwidth]{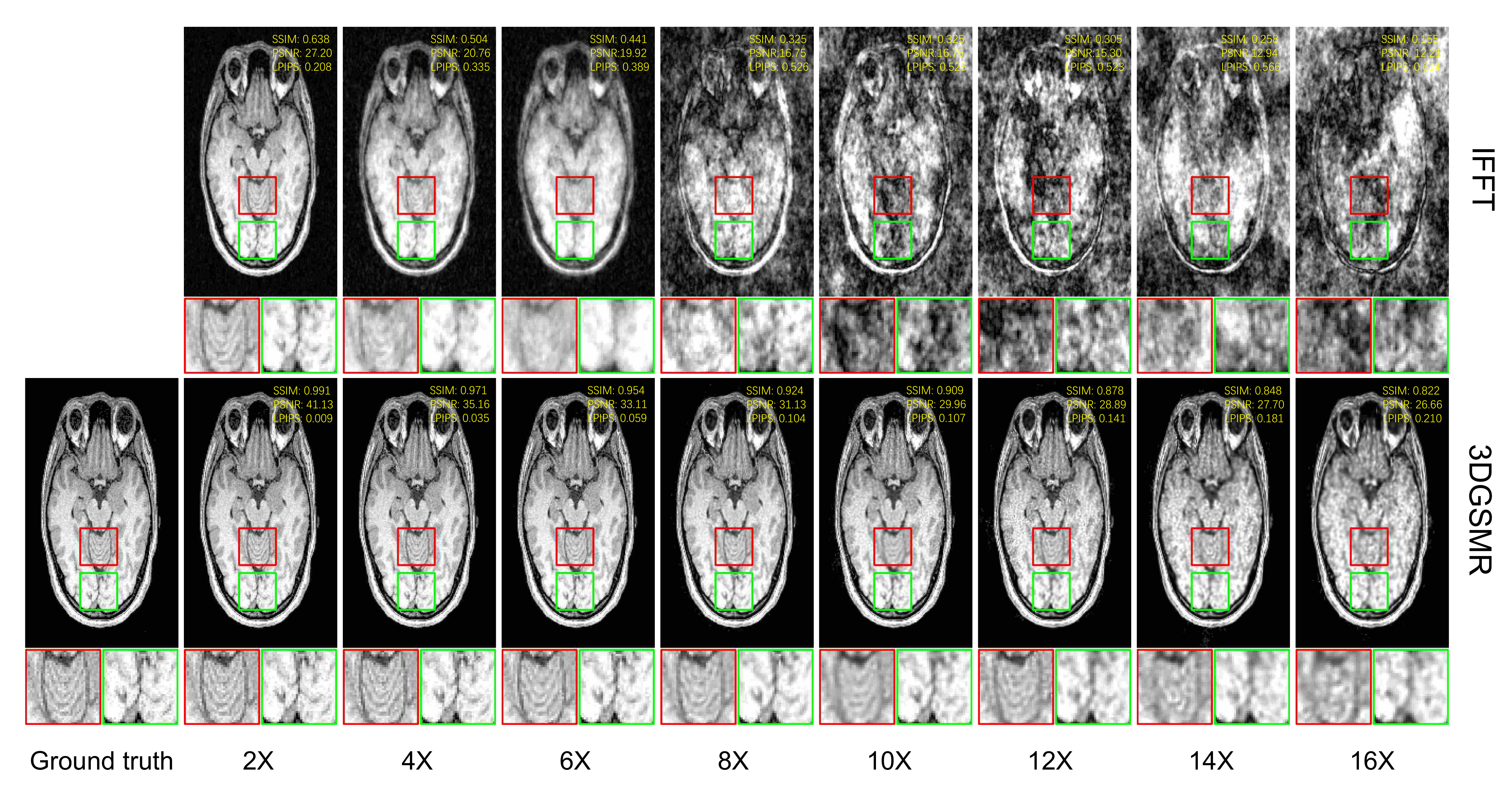}
  \caption{Showcase of the reconstruction results at acceleration factors of 2, 3, 4, 6, 8, 10, 12, 14, and 16 from one subject. The proposed framework demonstrates robust performance across different acceleration factors. As the acceleration factor continues to increase, the reconstructed images become more blurred, hindering the details in the images. Two regions in the selected coronal slice are zoomed in for better visual comparison.}
  \label{recon_results}
\end{figure*}

This section presents the reconstruction results across different acceleration factors. Figure \ref{recon_results} shows the reconstructions for different acceleration factors using the proposed framework as well as direct iFFT on one subject. Two regions in the selected coronal slice containing gray matters and white matters are magnified for further visual comparison. As the acceleration factor increases, the white matter and gray matter regions from the iFFT reconstruction become increasingly blurred and noisy, making it difficult to distinguish the details. While the proposed 3DGSMR framework is able to resolve these artifacts and reconstruct the details in the images.

In Figure \ref{recon_results}, quantitative results using SSIM, PSNR, and LPIPS are also present on each image. The iFFT results are used as the baseline to highlight the improvement of the proposed framework. The observed quantitative improvements underscore the effectiveness of 3DGSMR in isotropic-resolution 3D MRI reconstruction.

\subsection{Noise resistance characteristics of 3DGSMR}

\begin{figure}[htbp!]
\centering
   \includegraphics[width=\textwidth]{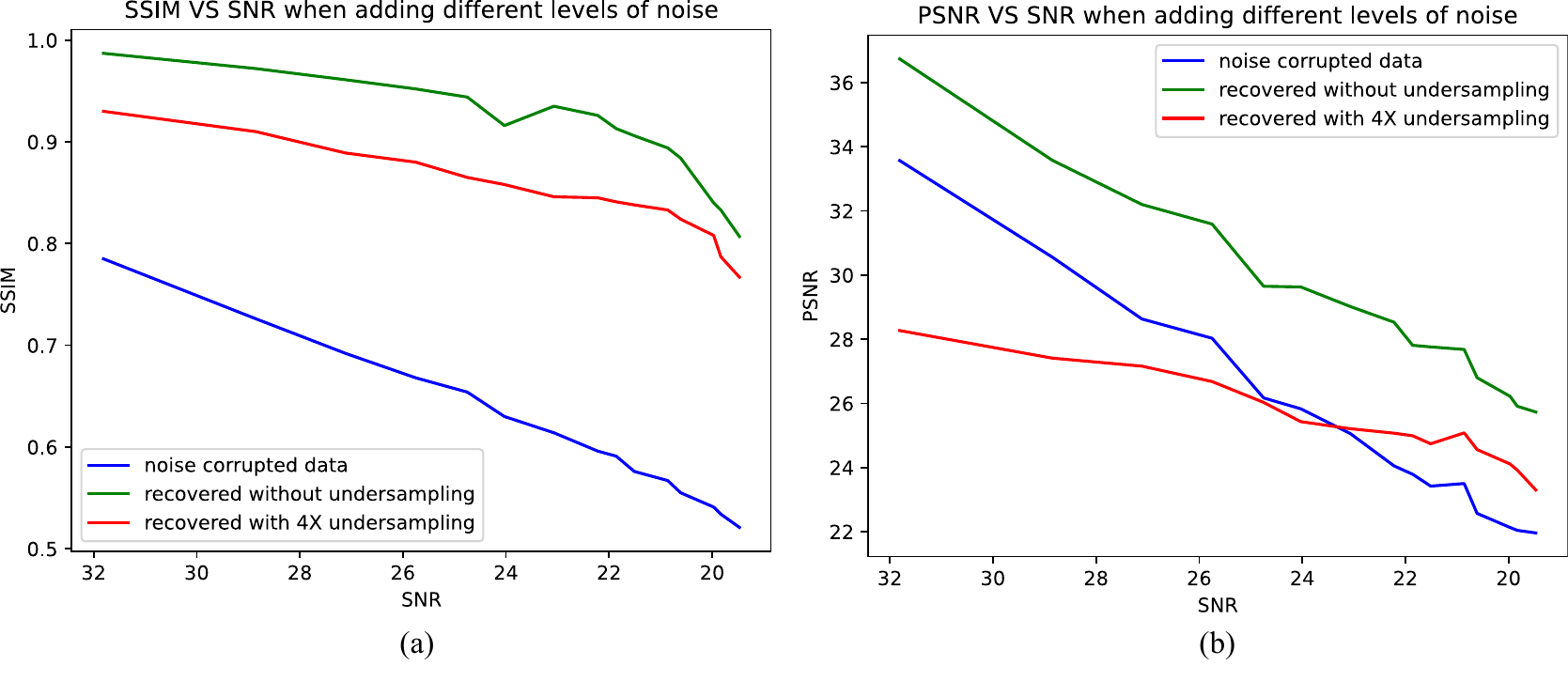}
  \caption{SSIM and PSNR against SNR when different level of noise is added into the k-space data. (a) SSIM (b) PSNR.}
  \label{metrics_SNR}
\end{figure}

\begin{figure*}[htbp!]
\centering
   \includegraphics[width=\textwidth]{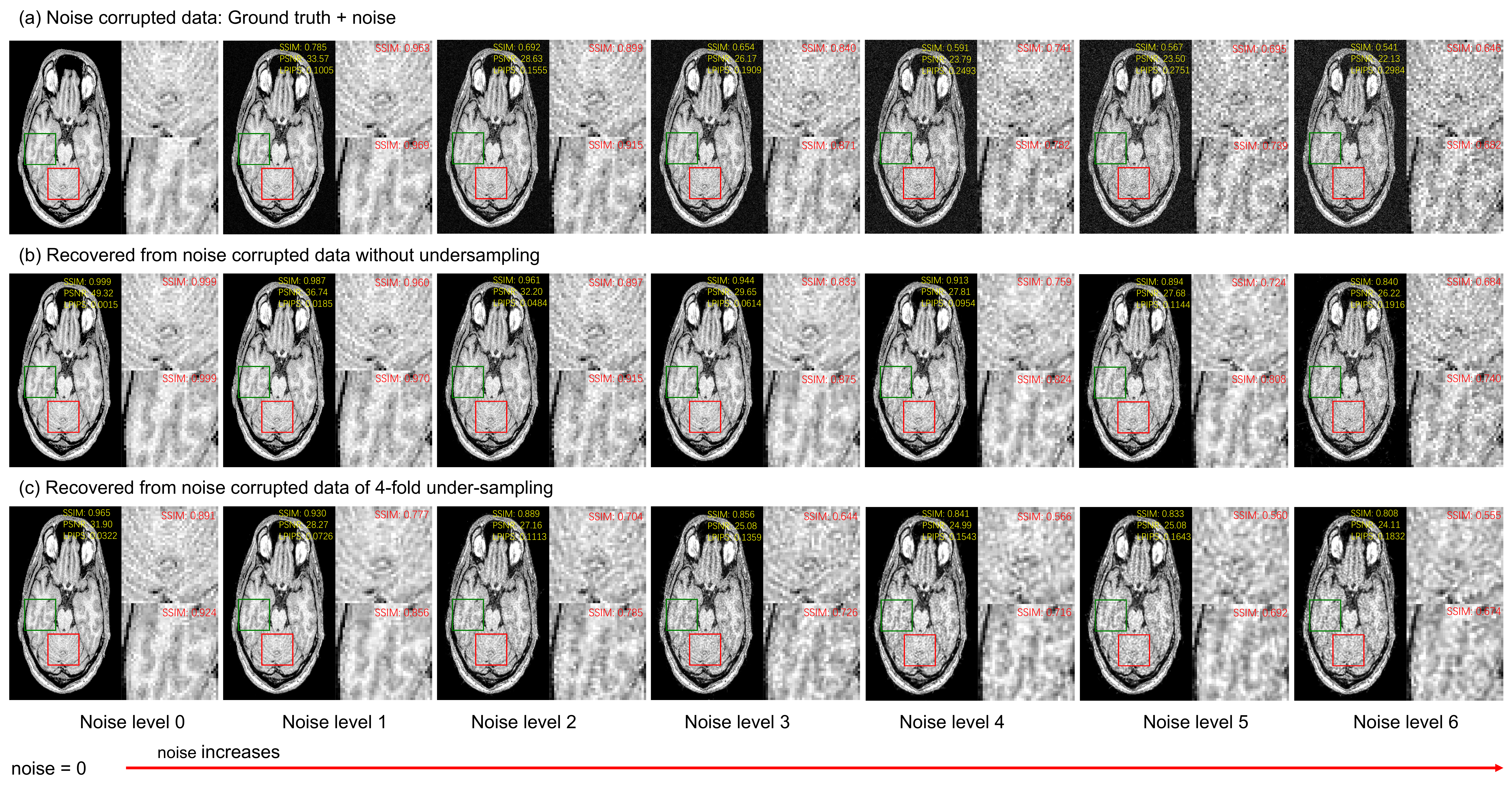}
  \caption{Comparison of reconstructed images obtained from varying levels of noise-corrupted data, evaluated using SSIM, PSNR, and LPIPS. The first row shows the noise-corrupted inputs, where the first image represents the noise-free ground truth. The second row presents the reconstructions from noise-corrupted data without undersampling, while the third row shows the reconstructions from 4-fold undersampled and noise-corrupted data. From left to right, the noise level gradually increases, leading to a progressive deterioration in reconstruction quality until it becomes visually unacceptable.}
  \label{noise_control}
\end{figure*}

It is observed that 3DGSMR could effectively reduce noise in the reconstruction. In this section, we present results under varying noise levels to demonstrate the noise-resistance capability of the proposed framework. Noise of different levels sampled from Gaussian distribution are manually added to the acquired k-space to simulate the noise scenario. The reconstruction results are shown in Figure \ref{noise_control}. The region marked in red is used to calculate the Signal-to-Noise ratio (SNR). Figure \ref{metrics_SNR} plots SSIM and PSNR as functions of noise level described in SNR. As shown in the figure, the proposed framework demonstrates noise‐reduction capability while maintaining fine structural details. When the noise level is relatively low, most noise in the image is effectively suppressed, and tissue textures are well preserved. As the noise level increases, the background of the MR images gradually becomes more contaminated, and the anatomical boundaries begin to blur. At extremely high noise levels—where nearly all voxels are corrupted and local neighborhoods appear uniformly noisy and blurred, the reconstructed results also deteriorate, and the fine textures can no longer be faithfully recovered.

\begin{figure}[htbp!]
\centering
   \includegraphics[width=1.0\textwidth]{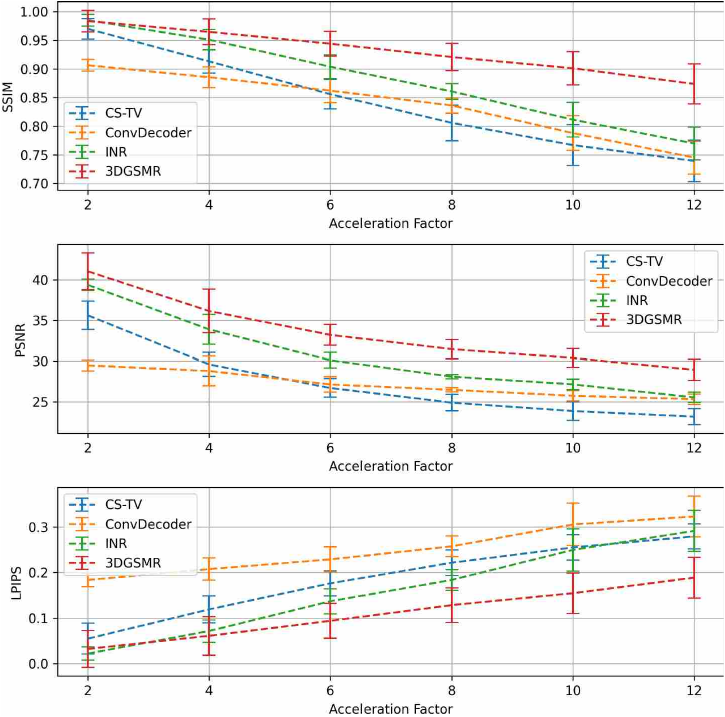}
  \caption{Qualitative and quantitative comparisons of the proposed 3DGSMR with CS-TV, ConvDecoder, and INR are presented for all the seven subjects, evaluating acceleration factors of 2, 4, 6, 8, 10, and 12 across SSIM, PSNR, and LPIPS metrics.}
  \label{fig:method_compare_curve}
\end{figure}

\subsection{Comparison study}
This section validates the effectiveness and robustness of the proposed method by exhibiting qualitative and visual comparison with other well-known and state-of-the-art isotropic resolution 3D MRI reconstruction methods introduced in the introduction section such as compressed-sensing with total variation regularization (CSTV) \cite{block2007undersampled}, ConvDecoder\cite{darestani2021accelerated}, and INR \cite{zhang2025low}. The later two are deep-learning based reconstruction methods that perform in a subject-specific fashion which allows reconstruction from undersampled k-space data without any priors and pre-training. Reconstruction results are evaluated and compared using metrics of SSIM, PSNR, and LPIPS agianst the groundtruth.

\begin{figure*}[htbp!]
\centering
   \includegraphics[width=\textwidth]{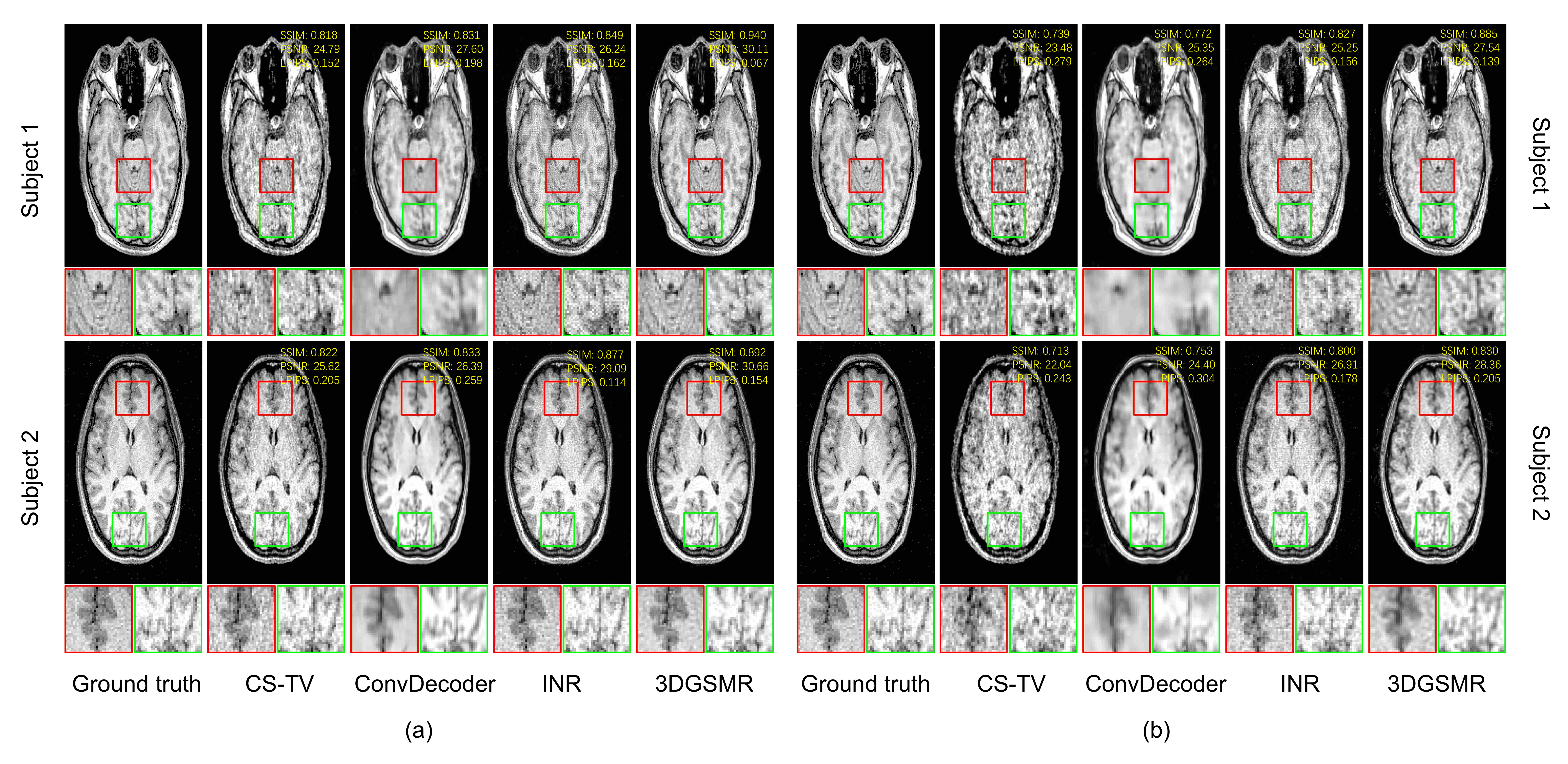}
  \caption{Visual comparison of the reconstruction results for CS-TV, ConvDecoder, INR, and proposed 3DGSMR on two subjects. (a) Visual comparison for acceleration factor of 8. (b) Visual comparison for acceleration factor of 10 with long-axis splitting. From the comparison, it can be found that the proposed scheme is able to recover the details with less distortions and artifacts.}
  \label{comp_sota}
\end{figure*}

\begin{figure*}[htbp!]
\centering
   \includegraphics[width=\textwidth]{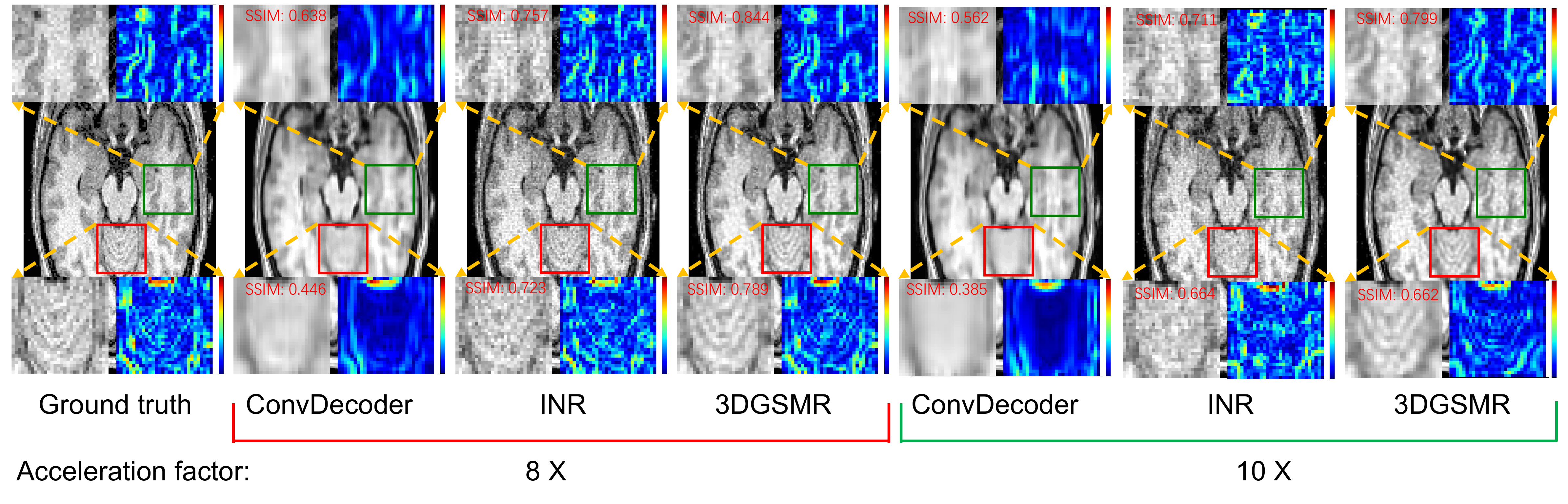}
  \caption{Zoomed-in visual comparison of the reconstruction results in edge-rich regions for ConvDecoder, INR, and proposed 3DGSMR on one subject with acceleration factor of 8 and 10. Two regions with rich details are selected for comparison. Corresponding patch-based SSIM is provided. Local gradients are computed and normalized to the same scale for comparison.}
  \label{gradient_comp}
\end{figure*}

In Figure \ref{comp_sota}, the visual comparison among different reconstruction methods are shown. Two axial slices from two different subjects together with two selected zoomed-in regions are present in the figure. In Figure \ref{gradient_comp}, another subject is selected, and two edge-rich patches are zoomed in for detailed analysis. Patch-based SSIM is reported to further substantiate the robustness of our model. Local gradient maps, computed using the Sobel operator, are also presented to demonstrate that 3DGSMR can effectively reconstruct and preserve fine structural details from undersampled acquisitions. The proposed method demonstrates superior performance visually as well as in SSIM and PSNR than the competing methods. Furthermore, the proposed method is able to restore the fine details in the images, closely resembling the ground truth. The improved image quality from the proposed reconstruction framework is also illustrated by the quantitative results, which are presented in Figure \ref{fig:method_compare_curve}.

\section{Discussion}
This study proposes an innovative approach for isotropic resolution 3D MRI reconstruction using 3DGS, leveraging its effectiveness to enable adaptive smoothing, interpolation, and precise spatial reconstruction from undersampled k-space data. This method has the potential to reduce the acquisition time for 3D MRI by highly undersample the k-space while maintaining reasonable image quality. Our experiments demonstrated that 3DGSMR can be effectively applied to isotropic resolution 3D MRI. 

Our model proved memory efficiency. 3DGS can be seen as a form of compression of the original voxelized representation of the images. By using 3D Gaussian points, it reduces the amount of memory required for storing detailed volume data, while simultaneously enhancing the continuous representation between voxels. The voxelization pipeline with CUDA parallelization smooths the transformation between 3DGS and voxelized representation, making it more efficient in terms of both memory usage and speed. In the absence of ground truth, the stopping point of the algorithm can be selected by monitoring the loss trajectory together with visual assessment of intermediate reconstructions. Additionally, the mean Sobel or Laplacian gradient within regions of interest may also serve as a quantitative proxy for the recovery of edge-rich detail.

The optimization scheme for 3DGS originated from \cite{zha2024r} proved its robustness in reconstruction at low-to-mid acceleration factors (such as 2X, 4X, 6X, and 8X). By eliminating cloning and replacing the original splitting method with the new long-axis splitting, and initializing from sparse Gaussian points, the proposed 3DGSMR model delivers impressive results with high acceleration factors (10X and 12X). It is also observed that the results from 3DGSMR were less noisy and maintained continuity in content,  demonstrating its ability to enhance image quality and improve visual appearance. We also demonstrated the model's ability to resist noise disturbance in k-space data. The denoising effect of the framework stems from its spatial smoothing of neighboring voxels. When noise corruption is sparse, the model can discriminate isolated noise from underlying anatomy. When noise corruption is spatially clustered, this distinction breaks down: the Gaussian representation may interpret the noise as genuine fine detail, drive the estimated kernels to extremely small variances, and converge to a noise-fitting solution. The time spent on the reconstruction can be considered efficient if the process is halted once an acceptable standard is reached. Typically, it takes 600 iterations for low-to-mid acceleration factors and 1,200 iterations for high acceleration factors that consume only few minutes to achieve clinically acceptable results. For acceleration factors greater than 12, the quality deteriorates significantly. For high acceleration factors, the original densification strategy introduces significant overlap in the densified Gaussian points, which amplifies the smoothing or blurring effect in the 3DGSMR reconstructions. This makes it harder for the model to converge towards small Gaussian points that capture fine details. The modified densification strategy that we proposed in this work uses only long-axis splitting without too much overlaps, allowing the Gaussian points to more easily split into smaller Gaussian points to recover the fined details. This approach facilitates the inference and recovery of the lost high-frequency components.

One of the advantages of 3DGSMR is that we could potentially apply the cutting-edged advancements of 3DGS into our proposed framework. Most recent works on 3DGS have focused on accelerating the optimization process or reducing aliasing artifacts \cite{lu2024scaffold, chen2025dashgaussian, lu2024turbo, hollein20243dgs}, while few have addressed memory efficiency or tackled the overfitting issues inherent in 3DGS \cite{liu2024compgs,park2025dropgaussian,morgenstern2024compact}. Zhao et al. introduced parallelized distributed training pipelines for 3DGS, significantly reducing training time and alleviating high GPU memory demands of 3DGS in Computer Vision \cite{zhao2024scaling}. Additionally, the original voxelizer implementation by Zha\cite{zha2024r} and the primitive model\cite{kerbl20233d} approximates 3DGS ellipsoids as spheres, using the longest semi-axis as the radius when computing Gaussian contributions to tiles. Currently, this bias has been solved by several works \cite{hanson2025speedy,wang2024adr,chen2025dashgaussian} to further reduce the GPU memory consumption and training time. We plan to investigate these advancements and introduce them into the proposed 3DGSMR framework in the future work.  

In this study, the 3D Gaussian points are mapped to grid points in k-space using the Fourier Transform. Each k-space sample encodes a specific spatial frequency and therefore influences every location in the image volume --- low frequencies govern coarse structure, while high frequencies capture fine detail. This global coupling means every frequency-domain grid point affects all Gaussian points in the image domain. In contrast, 3DGS for X-ray reconstruction is local along rays: a projection at a given angle updates only the Gaussian points intersected by that ray path. Moreover, the Gaussian distribution has a unique property where its Fourier transform remains a Gaussian distribution, meaning the frequency domain representation can be directly derived as combinations of Gaussian distributions from the 3DGS in image space. This approach is similar to the ''Splatting'' method in the original 3D Gaussian splatting \cite{kerbl20233d}. Consequently, the Fourier transform of a combination of 3DGS samples results in a corresponding combination of 3DGS samples in the frequency domain. This principle holds true even with non-uniform sampling, providing a promising strategy to circumvent the computational overhead typically associated with extensive FFT, interpolation, or other non-uniform FFT techniques. We plan to investigate this further in the future work.

The study in this work focused on 3D isotropic MRI reconstruction, and limited in-house data prevented us from implementing and comparing other cutting-edge methods, which requires data for pre-training. As a result, we only compared the model with subject-specific methods that do not rely on pre-training or priors.

The proposed model also inherits some limitations of the original 3DGS framework, such as needle-like artifacts under high acceleration factors, which could be viewed in Figure \ref{recon_results}. This suggests that a few Gaussian points may overfit without sufficient high resolution components in frequency domain for training. Furthermore, 3DGS faces challenges when processing low signal-to-noise ratio (SNR) images. The noise can be amplified during the Gaussian fitting process, leading to the introduction of artifacts or incorrect structures in the reconstructed image. The model may struggle to accurately differentiate between signal and noise when low SNR images are present, resulting in blurred details and inaccurate reconstructions, especially in high-frequency regions.

The convergence of 3DGS in medical image reconstruction remains an area with limited theoretical proof, particularly for clinical applications. While 3DGSMR shows promise in generating high-quality reconstructions for MRI, its ability to reliably converge to an ideal solution under various noise and data imperfections is not well-established. Furthermore, enhancing the model's interpretability, especially in understanding how Gaussian distributions influence the final reconstruction, is crucial for clinical adoption. Mathematical convergence proofs and clearer explainability would be essential to ensure the model's reliability and transparency in clinical decision-making.

\section{Conclusion}

In conclusion, this study has introduced and evaluated a novel approach, designated as 3DGSMR, which utilizes 3D Gaussian representations for the reconstruction of isotropic resolution 3D MRI from highly undersampled data. The 3DGSMR model has proven to be effective in managing the challenges associated with undersampled 3D MRI reconstruction, efficiently decomposing complex-valued MRI data using 3D Gaussian representations. Refinements were made in the densification strategies of the 3DGS, which have significantly accelerated the convergence process and improved the overall performance of the reconstructions at high acceleration factors. These advancements underscore the potential of 3DGSMR in enhancing the precision and efficiency of 3D MRI, thereby contributing valuable insights and tools to the field of medical imaging.

\bibliographystyle{IEEEbib}
\bibliography{refs}

\end{document}